\documentclass[10pt, a4paper, twocolumn]{article} 

\usepackage[svgnames]{xcolor}
\usepackage{amssymb}
\usepackage{pdfpages}

\usepackage[english]{babel} 

\usepackage{microtype} 

\usepackage{amsmath,amsfonts,amsthm} 

\usepackage[svgnames]{xcolor} 

\usepackage[small, labelfont=bf, up, textfont=it]{caption} 

\usepackage{booktabs} 

\usepackage{lastpage} 

\usepackage{graphicx} 

\usepackage{enumitem} 
\setlist{noitemsep} 

\usepackage{sectsty} 
\allsectionsfont{\usefont{OT1}{phv}{b}{n}} 

\usepackage{geometry} 

\usepackage[T1]{fontenc} 
\usepackage[utf8]{inputenc} 

\usepackage{XCharter} 

\usepackage{fancyhdr} 
\renewcommand{\sectionmark}[1]{\markboth{#1}{}} 

\fancypagestyle{firstpage}{ 
	\fancyhf{}
}

\newcommand{\authorstyle}[1]{{\large\usefont{OT1}{phv}{b}{n}\color{DarkRed}#1}} 

\newcommand{\institution}[1]{{\footnotesize\usefont{OT1}{phv}{m}{sl}\color{Black}#1}} 

\usepackage{titling} 

\newcommand{\HorRule}{\color{DarkGoldenrod}\rule{\linewidth}{1pt}} 

\pretitle{
	\vspace{-15mm} 
	\HorRule\vspace{10pt} 
	\fontsize{24}{28}\usefont{OT1}{phv}{b}{n}\selectfont 
	\color{DarkRed} 
}

\posttitle{\par\vskip 15pt} 

\preauthor{} 

\postauthor{ 
	\vspace{1mm} 
	\par\HorRule 
	\vspace{2mm} 
}

\usepackage{lettrine} 
\usepackage{fix-cm}	

\usepackage{xstring} 

\usepackage[super,nomove]{cite}

\title{Unusually High CO Abundance of the First Active Interstellar Comet} 

\author{
	\authorstyle{Cordiner, M. A.$^{1,2}$, Milam, S. N.$^1$, Biver, N.$^3$, Bockel{\'e}e-Morvan, D.$^3$, Roth, N. X.$^{1,4}$, Bergin, E. A.$^5$, Jehin, E.$^6$, Remijan, A. J.$^7$, Charnley, S. B.$^1$, Mumma, M. J.$^1$, Boissier, J.$^8$, Crovisier, J.$^3$, Paganini, L.$^9$,  Kuan, Y.-J.$^{10,11}$, Lis, D. C.$^{12}$} 
	\newline\newline 
	\textsuperscript{1}\institution{Solar System Exploration Division, NASA Goddard Space Flight Center, 8800 Greenbelt Road, Greenbelt, MD 20771, USA.}\\ 
	\textsuperscript{2}\institution{Department of Physics, Catholic University of America, Washington, DC 20064, USA.}\\ 
	\textsuperscript{3}\institution{LESIA, Observatoire de Paris, Universit{\'e} PSL, CNRS, Sorbonne Universit{\'e}, Universit{\'e} de Paris, 5 place Jules Janssen, 92195 Meudon, France.}\\ 
	\textsuperscript{4}\institution{Universities Space Research Association, Columbia, MD 21046, USA.}\\ 
	\textsuperscript{5}\institution{Department of Astronomy, University of Michigan, 311 West Hall, 1085 S. University Ave, Ann Arbor, MI 48109, USA.}\\ 
	\textsuperscript{6}\institution{STAR Institute, Universit{\'e} de Li{\`e}ge, All{\'e}e du 6 Aout, 19C, 4000 Li{\`e}ge, Belgium.}\\ 
	\textsuperscript{7}\institution{National Radio Astronomy Observatory, Charlottesville, VA 22903, USA.}\\ 
	\textsuperscript{8}\institution{IRAM, 300 Rue de la Piscine, 38406 Saint Martin d'Heres, France.}\\ 
	\textsuperscript{9}\institution{NASA Headquarters, Washington, DC, United States of America.}\\ 
	\textsuperscript{10}\institution{National Taiwan Normal University, Taipei 116, Taiwan, ROC.}\\ 
	\textsuperscript{11}\institution{Institute of Astronomy and Astrophysics, Academia
Sinica, Taipei 106, Taiwan, ROC.}\\
	\textsuperscript{12}\institution{Jet Propulsion Laboratory, California Institute of Technology, 4800 Oak Grove Drive, Pasadena, CA 91109, USA.}\\ 
}

\date{Published in Nature Astronomy, 20 April 2020} 

\begin{document}

\maketitle 

\thispagestyle{firstpage} 


{\bf Comets spend most of their lives at large distances from any star, during which time their interior compositions remain relatively unaltered. Cometary observations can therefore provide direct insight into the chemistry that occurred during their birth at the time of planet formation.\cite{mum11} To-date, there have been no confirmed observations of parent volatiles (gases released directly from the nucleus) of a comet from any planetary system other than our own. Here we present high-resolution, interferometric observations of 2I/Borisov, the first confirmed interstellar comet, obtained using the Atacama Large Millimeter/submillimeter Array (ALMA) on 15th-16th December 2019. Our observations reveal emission from hydrogen cyanide (HCN), and carbon monoxide (CO), coincident with the expected position of 2I/Borisov's nucleus, with production rates $Q({\rm HCN})=(7.0\pm1.1)\times10^{23}$\,s$^{-1}$ and $Q({\rm CO})=(4.4\pm0.7)\times10^{26}$\,s$^{-1}$. While the HCN abundance relative to water (0.06--0.16\%) appears similar to that of typical, previously observed comets in our Solar System, the abundance of CO (35--105\%) is among the highest observed in any comet within 2~au of the Sun.  This shows that 2I/Borisov must have formed in a relatively CO-rich environment --- probably beyond the CO ice-line in the very cold, outer regions of a distant protoplanetary accretion disk, as part of a population of small, icy bodies analogous to our Solar System's own proto-Kuiper Belt.}

During the last few decades, remote and \emph{in situ} measurements of volatiles and dust in comets belonging to our own Solar System have revealed a wealth of information on the chemical inventory and physical processes that occurred during the formation of our planets, 4.5 billion years ago\cite{mum11}. The chemical conditions prevalent during planet formation around other stars have become accessible for some nearby systems, thanks to advances in astronomical remote sensing methods\cite{cle18,pod19,ber19}, but this work is hindered by the extreme difficulty of observing ice and gas in the dense, opaque mid-planes of the disks where extrasolar planets form. Measurements of cometary compositions provide a unique method for probing the composition of the disk mid-plane, thus providing crucial input for theories concerning protoplanetary disk chemical evolution\cite{wal14,dro16}, and helping to improve our understanding of the ingredients available for forming planetary bodies and their atmospheres\cite{lod04,dod09,obe11}. Observations of the interstellar comet 2I/Borisov provide unique new insights into the chemistry that occurred during planet formation in another stellar system.

Comet 2I/Borisov (initially designated C/2019 Q4), was detected on 2019 August 30 by amateur astronomer G. Borisov, and later shown to be on a hyperbolic orbit (with eccentricity of 3.36), consistent with an interstellar origin in the direction of Cassiopeia\cite{bai20}. Although the apparition of interstellar objects (ISOs) was predicted for decades\cite{tor86}, 2I/Borisov is only the second confirmed ISO. The first --- 1I/'Oumuamua --- was discovered in October 2017 when it was already leaving the Solar System, making detailed studies difficult. Astronomical observations were able to partially constrain 'Oumuamua's shape and surface colors\cite{oum19}, and possible outgassing activity was implied by non-gravitational acceleration\cite{mic18}, but a lack of gas spectroscopic detections made a cometary classification for 1I/'Oumuamua uncertain. 

The presence of a gas and dust coma surrounding 2I/Borisov, on the other hand, has been confirmed by numerous observations, including the detection of cyanide (CN) and hydroxyl (OH) emission in the near-UV and optical\cite{fit19,opi19,bod20}. The observations reported to-date have revealed a CN/H$_2$O ratio and optical dust colours consistent with typical Solar System comets, while the C$_2$ abundance may be somewhat depleted\cite{mck20,jew19}. The OH radical is believed to be a product of water photodissociation in the coma, but the origin of 2I/Borisov's CN is less certain, as cometary CN can be produced from the photodissociation of HCN and other nitriles, and from the degradation of organic-rich dust grains\cite{fra05}. In contrast, observations of rotational emission in the microwave and sub-mm part of the spectrum\cite{boc18} can reveal parent volatiles --- sublimating gases that have been stored (as ices) inside 2I/Borisov's nucleus, for the duration of its interstellar journey --- and therefore provide a more direct probe of its chemical composition. 

We used the Atacama Large Millimeter/submillimeter Array (ALMA) to obtain spectra and images of 2I/Borisov in the frequency range 342.9-355.6 GHz (0.84-0.87~mm), during its passage through the inner Solar System. The comet reached perihelion on 2019 December 8th, and our observations were carried out over two sessions on 2019 December 15th and 16th, at a heliocentric distance $r_H=2.01$~au and a geocentric distance $\Delta=1.96$~au. Further details of the observations and data analysis procedures are presented in the Methods section.

The targeted emission lines of HCN ($J=4-3$) and CO ($J=3-2$) were both detected at the center of the ALMA field of view (see Figure 1). Spectrally integrated fluxes are given in Table 1, and correspond to a statistical significance of $5.2\sigma$ for HCN and $5.9\sigma$ for CO. Our HCN and CO detections are robust, but the detailed coma morphology in Figure 1 is largely hidden by noise; HCN shows a central peak, slightly extended in the north-south direction, while an extended flux component is detectable for CO, at distances up to 4000~km from the nucleus. Our observations also targeted CS and CH$_3$OH, but these molecules were not detected.

\begin{figure}[!ht]
\centering
\hspace*{-2.5mm}
\includegraphics[width=1.05\columnwidth]{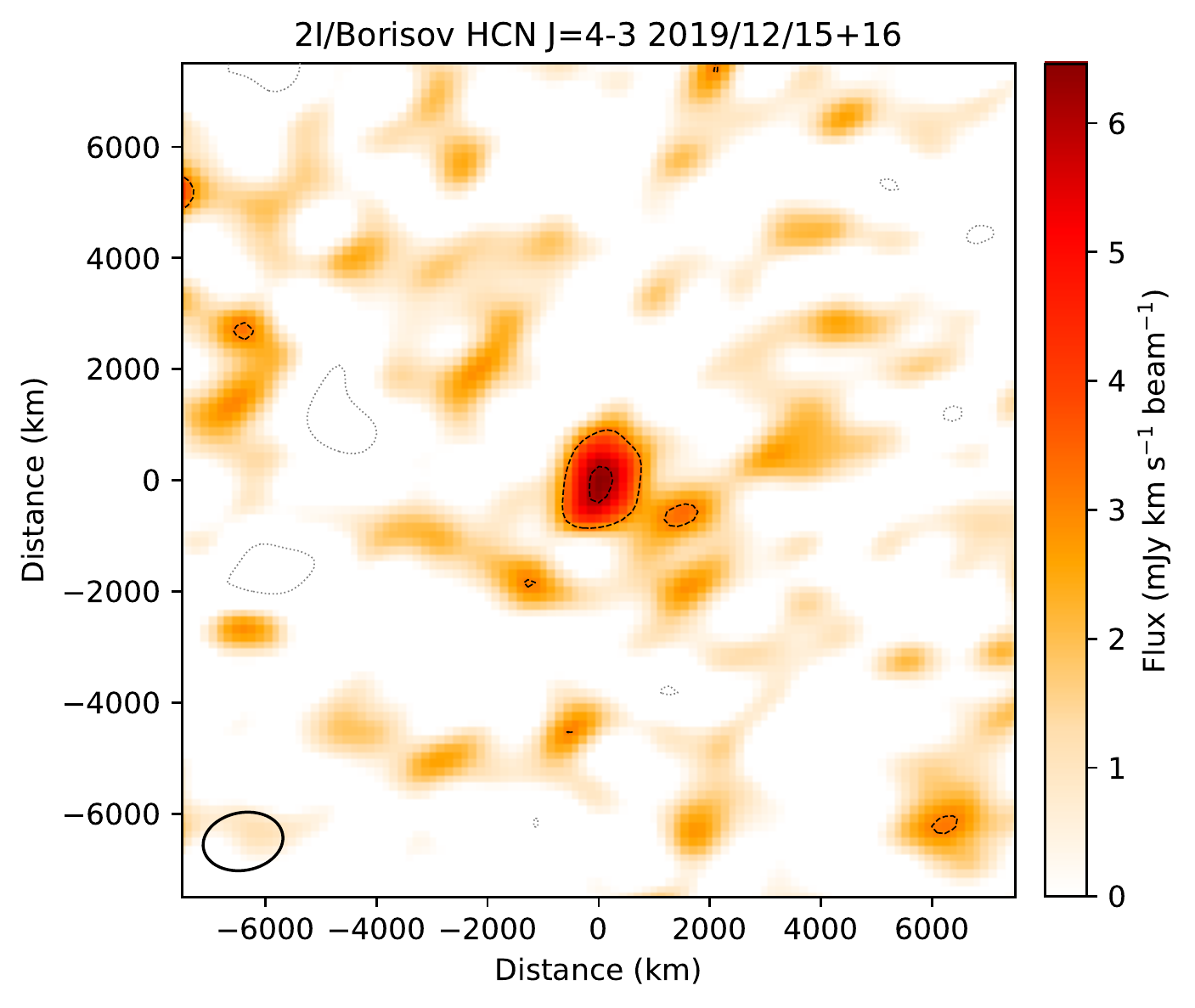}\\
\hspace*{-2.5mm}
\includegraphics[width=1.05\columnwidth]{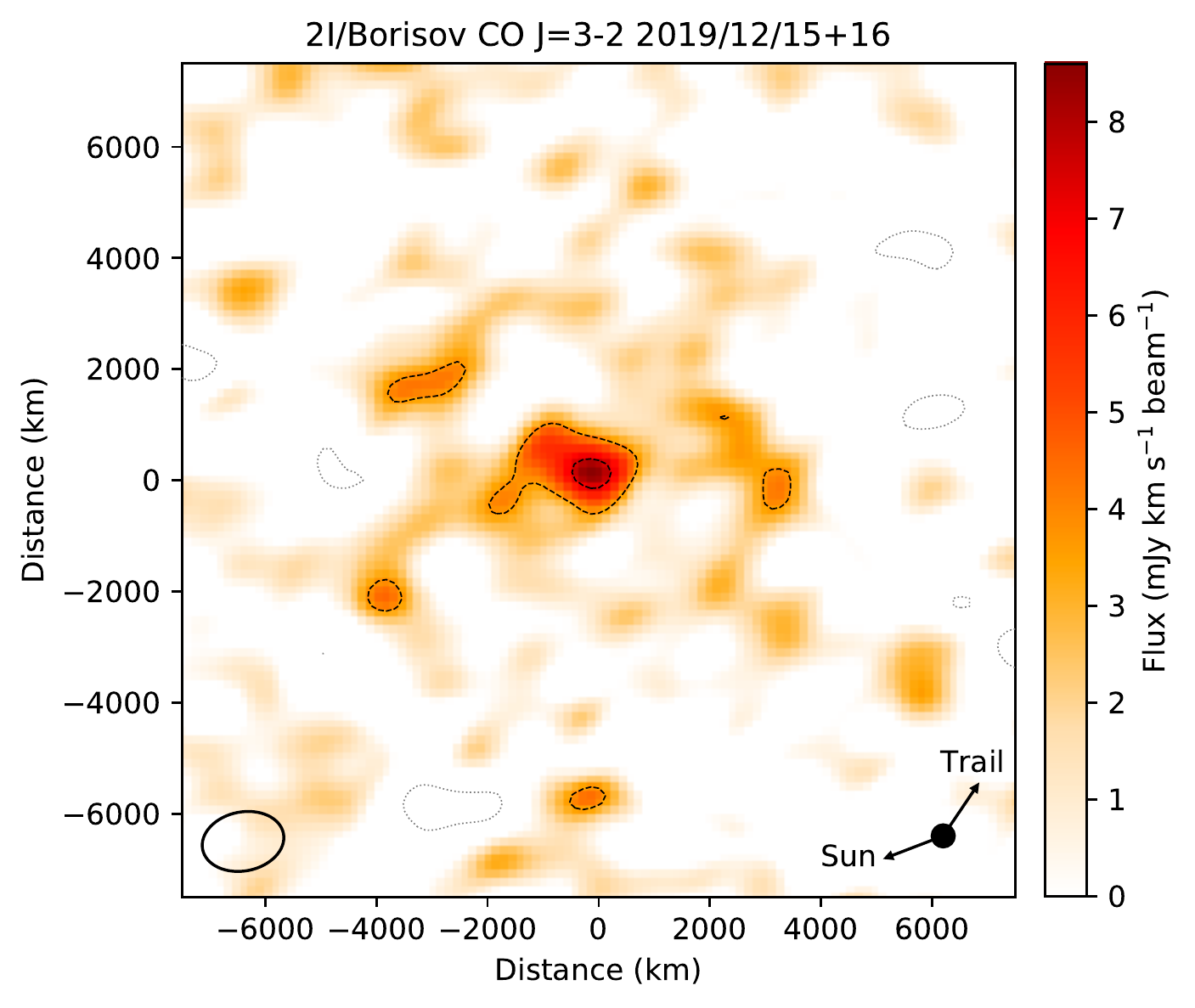}
\caption{Spectrally integrated flux maps of HCN ($J=4-3$) and CO ($J=3-2$) emission from interstellar comet 2I/Borisov. Positive flux contours are shown with dashed lines, and negative contours are dotted, with a contour spacing equal to 2.5 times the RMS noise. FWHM of the spatial resolution element ($1.0''\times0.7''\approx1500\times1000$~km) is indicated lower-left. Sky-projected solar and orbital trail vectors are shown lower-right. Coordinate axes are aligned with the equatorial system, with the origin at the HCN peak.}
\end{figure}

\begin{figure}[!ht]
\centering
\includegraphics[width=\columnwidth]{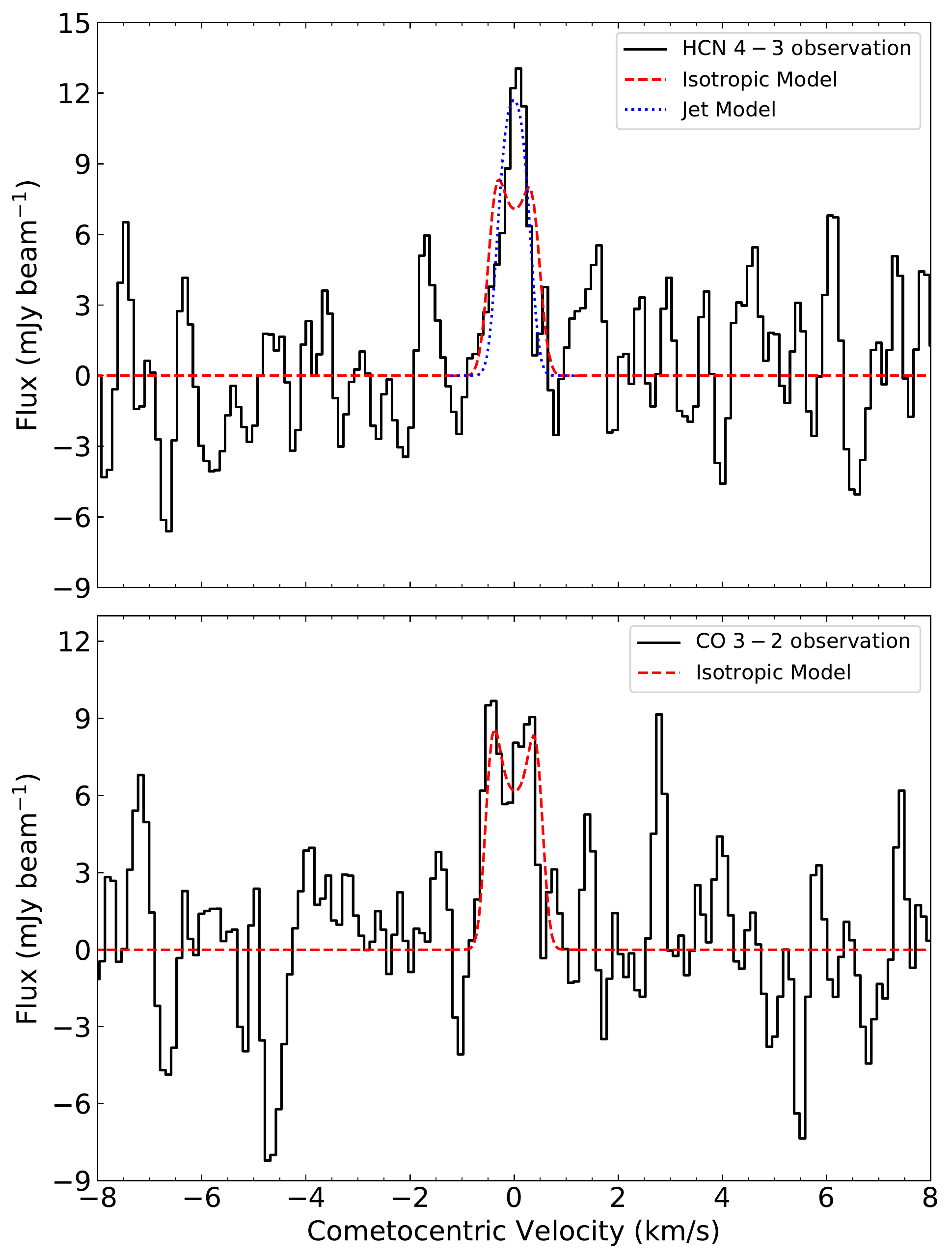}
\caption{ALMA interferometric spectra of HCN (top) and CO (bottom) towards 2I/Borisov, extracted at the respective emission peaks of each species (see Figure 1). Best-fitting radiative transfer models are overlaid (see Methods), including isotropic outgassing (at $v_{out}=0.47$~km\,s$^{-1}$),  with an alternative conical jet model shown for HCN, for which the outflowing gas is confined within a 96$^{\circ}$ cone, with its apex at the nucleus and axis in the plane of the sky. The spectral resolution is 0.21~km\,s$^{-1}$ (channel spacing = 0.11~km\,s$^{-1}$).}
\end{figure}

\begin{figure}
\centering
\includegraphics[width=\columnwidth]{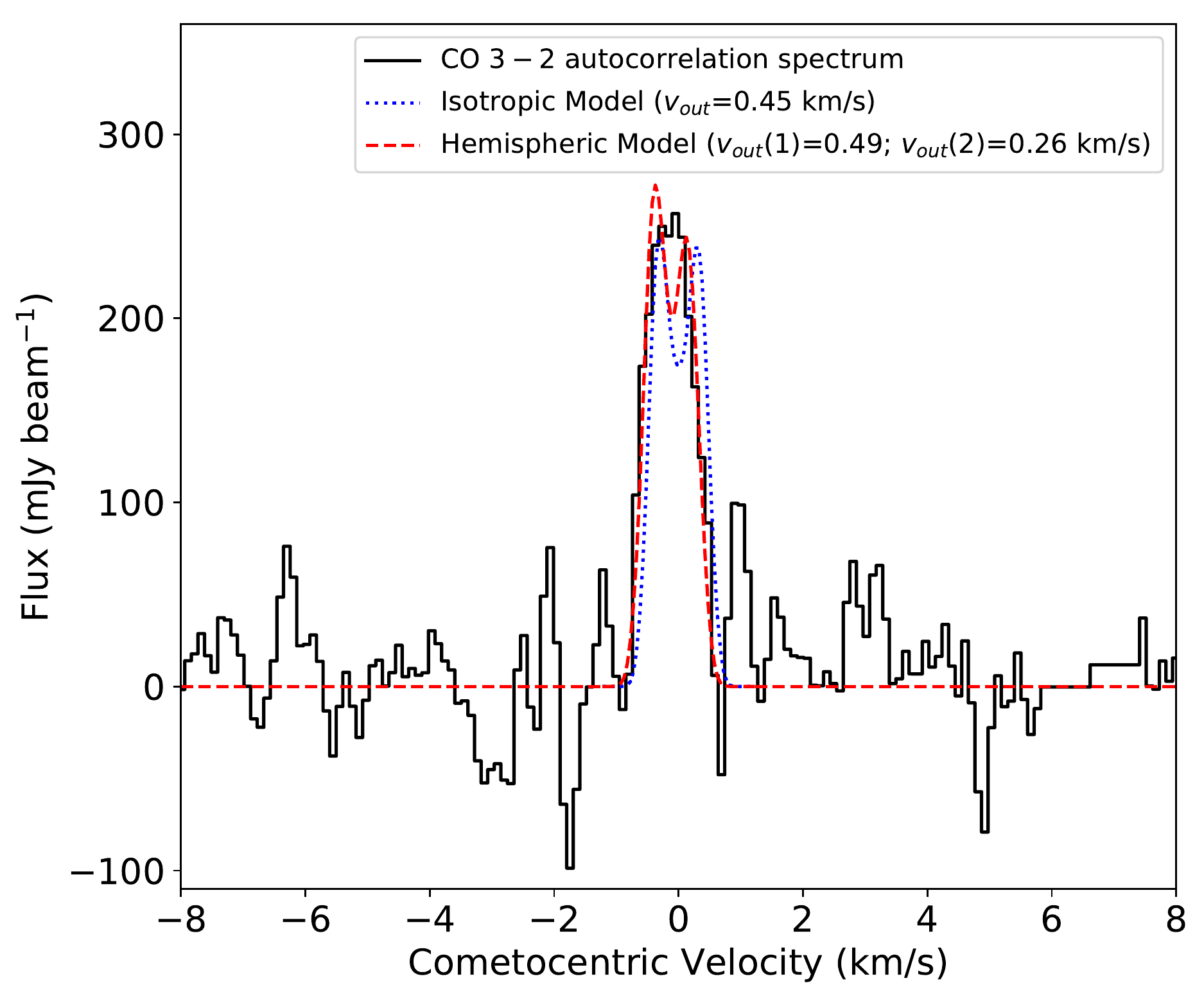}
\caption{ALMA autocorrelation (total power) spectrum of 2I/Borisov's CO, averaged over all 42 antennas (observed UT 2019-12-15). Two best-fitting models are overlaid, assuming isotropic outgassing (dotted blue line) and asymmetric outgassing along the sun-comet vector (dashed red line). The integrated line intensity is $260\pm15$~ mJy\, km\,s$^{-1}$\, beam$^{-1}$ (equivalent to $9.3\pm0.5$~mK\,km\,s$^{-1}$ on the main beam brightness temperature scale).}
\end{figure}

Molecular production rates ($Q$) were obtained by performing nonlinear least-squares fits to the extracted spectra, shown in Figure 2, using a three-dimensional radiative transfer model (see Methods). This resulted in $Q({\rm CO})=(4.4\pm0.7)\times10^{26}$\,s$^{-1}$ and $Q({\rm HCN})=(7.0\pm1.1)\times10^{23}$\,s$^{-1}$, assuming CO and HCN were well mixed, with a constant kinetic temperature of $T_{kin}=50$~K. For CO, the spectral line profile (with FWHM = $1.1\pm0.1$~km\,s$^{-1}$) is consistent with an isotropically expanding coma, with best-fitting outflow velocity $v_{out}=0.47\pm0.04$~km\,s$^{-1}$.  HCN exhibits a narrower spectral line shape (with FWHM = $0.57\pm0.09$~km\,s$^{-1}$), that could arise as a result of localised HCN outgassing in the form of a jet-like stream primarily confined to the plane of the sky, with correspondingly narrower radial velocity dispersion. However, the signal-to-noise is too low to conclusively distinguish this scenario from isotropic outgassing.  Observations of near-UV CN fluorescence using the TRAPPIST-South telescope on UT 2019-12-15 gave $Q({\rm CN})=(9.5\pm1.8)\times10^{23}$\,s$^{-1}$, which is consistent (within uncertainties) with our derived HCN production rate.

A clearer view of the CO $J=3-2$ emission line is shown in Figure 3, which was obtained by averaging only the autocorrelation (total power) spectra obtained from each of the individual ALMA antennas\cite{cor19}. These autocorrelation data are more sensitive to 2I/Borisov's CO emission than the interferometric data (in Figures 1 and 2), because they include flux from angular scales $\gtrsim10''$, which was filtered out by the interferometer due to a lack of very short baselines in the telescope array.  Modeling the autocorrelation data assuming isotropic outgassing from the nucleus results in a CO production rate of $(5.0\pm0.5)\times10^{26}$~s$^{-1}$ (assuming an amplitude calibration error of $\pm10$\%). This is consistent with the value obtained from the interferometric data (within errors), adding confidence to our CO measurements. HCN was not detected in the autocorrelation data, which is consistent with a faster drop in HCN flux with radius, due to more rapid radiative cooling (depopulation) of the HCN $J=4$ level than for the $J=3$ level of CO. This is a result of HCN's larger dipole moment, combined with a lack of thermalizing collisions in 2I/Borisov's relatively low-density coma.

Assuming 2I/Borisov's water production rate on December 15th-16th was within the range 4.9-10.7$\times10^{26}$~s$^{-1}$ observed by Xing et al.\cite{bod20} (see Methods), we find that the HCN abundance (relative to H$_2$O) was 0.06\%-0.16\% and the CO abundance was 35\%-105\%. This demonstrates a relatively unusual composition for 2I/Borisov's coma: HCN is close to the average abundance of 0.12\% observed previously in Solar System comets\cite{boc18}, while the CO abundance is much higher than the average value of 4\%\cite{pag14a}. Even Oort Cloud comets previously designated as CO-rich only have CO/H$_2$O ratios in the range 10\% to 24\%\cite{pag14a} (see also Supplementary Table 1), with the exception of the recently discovered, peculiar Oort Cloud comet C/2016 R2 (PanSTARRS), which had CO/H$_2$O = 30,800\%\cite{mck19}.

\begin{figure}[!ht]
\centering
\includegraphics[width=\columnwidth]{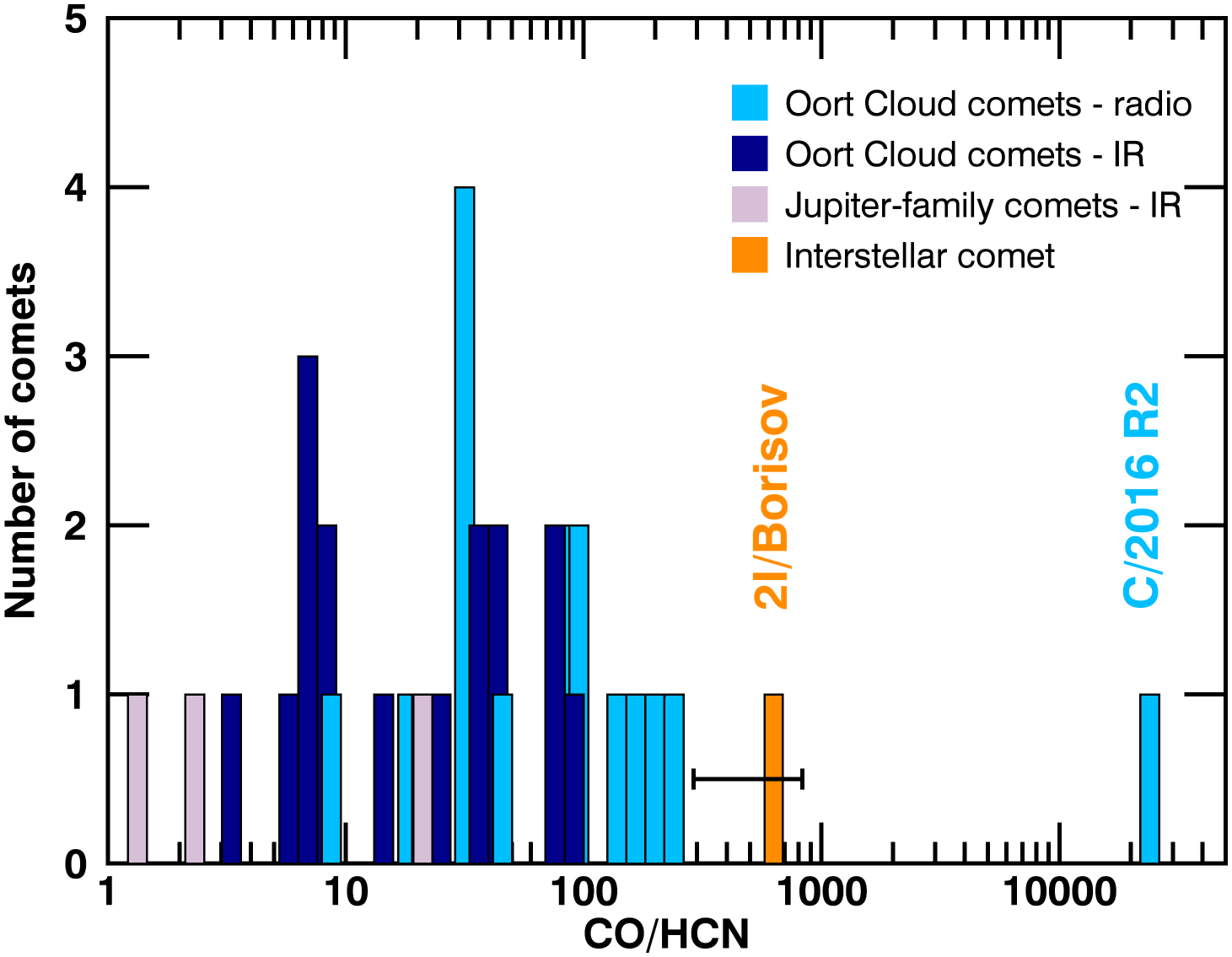}
\caption{Histogram showing previously-published CO/HCN mixing ratios observed in Solar System comets\cite{del16,boc18}. 24 Oort Cloud comets and 3 Jupiter-family comets are shown. The unusually high CO/HCN ratio of 2I/Borisov ($630^{+200}_{-340}$) is highlighted, along with the chemically peculiar outlier C/2016 R2 (PanSTARRS) (which had CO/HCN = 26,400)\cite{biv18} (horizontal black bar indicates the uncertainty range of our analysis). An annotated version of this plot is given in Supplementary Figure 1.}
\end{figure}

\begin{table*}
\centering
\caption{Observed Molecular Lines, Peak Fluxes (per beam) and Production Rates }
\vspace*{-1mm}
\begin{tabular}{lcccr}
\hline\hline
Molecule & Transition & Frequency & Flux & Production Rate \\
         &            & (GHz)     & (mJy\,km\,s$^{-1}$\, bm$^{-1}$) & (mol.\,s$^{-1}$)\\
\hline  
CO       & $J=3-2$              & 345.796 & $8.6\pm1.5$ & $(4.4\pm0.7)\times10^{26}$ \\
HCN      & $J=4-3$              & 354.505 & $6.5\pm1.2$ & $(7.0\pm1.1)\times10^{23}$ \\
CS       & $J=7-6$              & 342.883 & $<5.2$      & $<6.5\times10^{23}$ \\
CH$_3$OH & $J_K=13_0 - 12_1 A^+$  & 355.603 & $<4.4$    & $<4.4\times10^{26}$ \\
\hline
\end{tabular}
\parbox{0.8\textwidth}{\vspace*{1.25mm} Table Footnotes --- Derived production rates from the ALMA interferometric data, assuming a kinetic temperature of 50~K (see Methods).}
\end{table*}

The CO/HCN mixing ratio of 2I/Borisov is shown for comparison with Solar System comets in Figure 4. Accounting for additional uncertainties in the HCN production rate due to possible variations in the coma kinetic temperature and electron abundance (see Methods), we find that the maximum error margin for CO/HCN is $630^{+200}_{-340}$. This CO/HCN ratio is still higher than seen before in any comet (see Supplementary Table 1), apart from C/2016 R2, which had CO/HCN = 26,400 (at $r_H=2.8$ au)\cite{biv18}. This provokes the idea that the interstellar comet 2I/Borisov may be chemically distinct from those typically found in our own Solar System. An important point to consider regarding this interpretation, however, is the degree to which the apparent CO-richness of 2I/Borisov's coma reflects its intrinsic nucleus composition, or instead, whether the differential sublimation rates as a function of temperature for CO, HCN and H$_2$O could play a role.

Supplementary Table 1 provides literature data for CO and HCN in five comets observed at $r_H=1.7$--$2.4$~au, including several CO-rich comets. The spread of values for CO/H$_2$O ($0.04\pm0.01$ to $0.32\pm0.05$) and CO/HCN ($36\pm13$ to $200\pm22$) reveals a considerable diversity in the relative CO activity level across this distance range, but confirms 2I/Borisov as one of the most CO-rich comets ever observed. After 2I/Borisov, the next highest CO/HCN ratio ($238\pm21$) was for C/2006 W3 (Christensen), at a relatively large heliocentric distance $r_H=3.3$~au\cite{boc10}. Cometary CO ice sublimates at a lower temperature than H$_2$O and HCN, so the CO/H$_2$O and CO/HCN production rate ratios should increase with heliocentric distance\cite{wie18}. Although the 95-150~K sublimation temperatures ($T_{sub}$) of HCN and H$_2$O are expected to be reached on the illuminated surfaces of comets within $r_H\sim3$~au\cite{wom17}, sub-surface ices (inside 2I/Borisov's nucleus) would be cooler, potentially allowing internal CO to sublimate while H$_2$O remains frozen. However, in a sample of 17 comets with CO$_2$ measured by the AKARI satellite, a relative reduction in the H$_2$O production rate was only found to occur at distances $r_H>2.5$~au\cite{oot12}. Our measured CO/HCN ratio in 2I/Borisov at 2.0 au is therefore likely to be representative of the interior ice mixing ratio.  

There remains a possibility that 2I/Borisov's enhanced CO/H$_2$O ratio arises as a consequence of thermally-insulating exterior layers, which could inhibit heat penetration, leading to a drop in the overall water production rate without significantly affecting that of the more volatile CO. The presence of such a low-thermal-conductivity crust was theorized as a possible explanation for the complete absence of observable outgassing from 1I/'Oumuamua\cite{fit18}, and could plausibly result from volatile depletion in 2I/Borisov's surface as a result of exposure to cosmic rays or other heat sources during its long interstellar journey. Surface modifications are also possible as a result of accretion of interstellar matter, or sputtering/abrasion by collision with particles, as the comet passed through the interstellar medium\cite{ste90}. A more complete investigation of the importance of such effects is warranted.

Given the (lack of) information on 2I/Borisov's internal physical structure, we now consider the implications of an intrinsically elevated CO abundance. CO-rich comets form in protoplanetary disks beyond, or in the vicinity of, the CO sublimation front (ice-line)\cite{ber18,eis18}, and these objects should have the full complement of CO and H$_2$O supplied from their parent interstellar cloud.  The CO ice line in our protosolar nebula has been estimated to be located in the outer part of the cometary formation zone, between about 17-35 au\cite{qi13,mor19}. Currently, the outer edge of the Solar System (as defined by a drop-off in the Kuiper Belt density), is close to 50 au\cite{gla08}.  In other planetary systems, the CO ice-line has been identified at similar or greater distances from the star\cite{qi13,zha19,qi19}, with a radius depending on the luminosity of the young star.  Observations of disks during the initial stages of planetary assembly find that their outer radii extend 5-200 au\cite{and20}.  Thus, there exists a population of young planet-forming regions with proto-Kuiper belts that have extensive CO-ice dominated zones, in which CO-rich comets may be forming.  Furthermore, ALMA has recently inferred the presence of large, Neptune to Saturnian sized objects at these distances\cite{zha18}.  These proto-planets would disturb the natal Kuiper-belt analog population and some comets with perturbed orbits would be launched on escape trajectories into interstellar space, occasionally ending up on hyperbolic orbits passing through our inner Solar System. Scattering events involving gravitational encounters with passing stars, or the members of a multiple star system, can also be considered as plausible ejection mechanisms.

Models for the formation and composition of cometary ices suggest significant chemical evolution within the protoplanetary disk close to the ice-lines of key volatiles such as H$_2$O, CO, and CO$_2$ \cite{obe11,eis18}.  It has long been theorized that evaporated ices inside the vapor region of a given ice-line would diffuse outwards (away from the star) and potentially enhance the ice-content of materials at larger radii\cite{ste88}. The CO-rich composition of comet 2I/Borisov therefore hints at formation in close vicinity to the CO ice-line in its natal protoplanetary disk.

Radiogenic heating of larger Kuiper-belt objects/small planets can enable the sublimation/diffusion of interior CO ice, which may then re-freeze in the cooler regions closer to the object's surface, thus leading to CO-enriched layers \cite{des01}. This leads to the final mechanism we envisage as plausible for the generation of CO-rich planetesimals --- through fragmentation (and subsequent ejection) of a chemically-differentiated parent body --- as was considered previously for C/2016 R2 (PanSTARRS)\cite{biv18,mck19}.

The detection of HCN and CO emission from the first confirmed, active interstellar comet represents a major breakthrough, providing unprecedented insights into the chemistry of a planetary system other than our own, and proving that simple (carbon, nitrogen and oxygen-bearing) molecular ices were abundant in 2I/Borisov's natal environment. While it is difficult to be sure that the observed coma abundances represent the bulk composition of volatiles in 2I/Borisov's nucleus, comparison with other comets observed at similar heliocentric distances reveals an unusual, CO-rich composition for this interstellar object, showing that it probably originated from the icy, outer regions of a distant protoplanetary disk, and maintained a very low interior temperature ($\lesssim25$~K) for the duration of its interstellar journey.  Considering the broader population of interstellar comets, our ALMA observations show that we should be prepared for the unexpected as we pursue future studies of this new class of astronomical object.

\section*{Methods}

\subsection*{Observations}

ALMA observations were performed as part of program 2019.01008.T on UT 2019-12-15 8:50--12:03 and UT 2019-12-16 11:37--12:56, with a total observing time of 1:11 on-source. The correlator was configured to observe rotational transitions of CS (342.883~GHz), CO (345.796~GHz), HCN (354.505~GHz) and CH$_3$OH (355.603~GHz), with spectral resolutions of 282~kHz, 244~kHz, 244~kHz, and 488~kHz, respectively. There were 42 active antennas in the array (configuration C43-1), spanning baselines 15-317~m, resulting in a spatial resolution (elliptical beam FWHM) of $1.02''\times0.73''$ at 355 GHz. Flux and phase calibration were performed using the quasar J1209-2406. The observing sequence consisted of 30~s on the phase calibrator, 2~min on the comet, and 2~min on a sky reference position $2'$ North of the comet (for the purpose of calibrating the ALMA autocorrelation data), repeated until the end of the observing period.  The array phase-center was updated in real-time to track 2I/Borisov's position on the sky, following JPL ephemeris solution \#48 (generated on 2019-12-05).

Data were flagged and calibrated with CASA version 5.6\cite{jae08} using the automated pipeline scripts supplied by the Joint ALMA Observatory. Deconvolution, Doppler correction and imaging were performed using the {\tt tclean} (Clark) algorithm, with natural weighting, a flux threshold of twice the RMS noise per channel ($\sigma_c$) in each image, and a $6''$-diameter circular mask about the phase center. The resulting, combined image cubes are in the rest frame of the comet, with $\sigma_c=2.6$~mJy for CO and 2.9~mJy for HCN.

The spectral cubes were integrated over the cometocentric velocity range $-0.3$ to $0.3$~km\,s$^{-1}$ for HCN and $-0.6$ to $0.6$~km\,s$^{-1}$ for CO to produce the maps shown in Figure 1. Spectra for CO and HCN were then extracted at the respective flux peaks for each species (both of which were within $0.3''$ of the phase center). For CS and CH$_3$OH (not detected), the spectra were extracted at the HCN peak ($0.2''$ East of center). Integrated fluxes (per beam) for each species are given in Table 1. No continuum emission was detected; the RMS noise on our combined continuum image resulted in a $3\sigma$ upper limit to the dust/nucleus thermal emission of 0.11~mJy\,beam$^{-1}$.

Autocorrelation (total power) spectra were recorded using each of the forty-two 12~m ALMA antennas, simultaneously with the interferometric (cross correlation) observations. These were calibrated and combined following the method of Cordiner et al.\cite{cor19}, using the interleaved sky reference position to subtract thermal contributions from the sky, receiver and telescope optics from each scan. Median system temperatures were 120~K at 346~GHz. The zenith sky opacity was 0.11 and the single-dish beam FWHM was $16.8''$. The data were corrected for a beam efficiency $\eta_{\rm MB}=0.72$, which was derived from the aperture efficiency ($\eta_A=0.63$ at 345~GHz)\cite{alm19} using the standard relationship\cite{dow89} $\eta_{\rm MB}=0.8899k^2\eta_A$, with k=1.13 for the ALMA 12-m antennas. The calibrated autocorrelation spectra were then Doppler shifted from the topocentric frame to the comet's rest velocity. A spectral baseline offset was subtracted using low-order polynomial fits to the line-free channels. A narrow residual feature due to incomplete telluric CO cancellation (at 6.6~km\,s$^{-1}$ in the comet's rest frame) was removed by linear interpolation.

\subsection*{Radiative Transfer Modeling}

The observed spectra were modeled with a nonlinear least-squares fitting procedure, using a modified version of the LIne Modeling Engine (LIME)\cite{bri10}, adapted for cometary comae as described by Cordiner et al.\cite{cor19}. This three-dimensional, non-local-thermodynamic-equilibrium (NLTE) radiative transfer model incorporates a calculation of the rotational energy level populations resulting from collisions with gas-phase molecules and electrons, as well as pumping by solar radiation. Collisional rates for HCN with H$_2$O and CO were taken from the H$_2$ collisional rates in the Leiden Atomic and Molecular Database \cite{sch05}. The impact of this assumption was tested by scaling the H$_2$ collision rates by a factor of 0.1-10, but the resulting HCN production rates changed by only 2\%. Radiative pumping and photodissociation rates\cite{hue15} were appropriately scaled for the comet's heliocentric distance ($r_H=2.01$~au). Due to the long ($\sim10^4$~s) lifetime of the CO $J=3$ level against radiative transitions, LTE was found to be appropriate for modeling our observations of this molecule, with the benefit of improved computational efficiency.

The comet's H$_2$O production rate was obtained from UV observations of OH using the SWIFT satellite\cite{bod20}. A linear interpolation of the observed values on December 1st ($(10.7\pm1.2)\times10^{26}$~s$^{-1}$) and December 21st ($(4.9\pm0.9)\times10^{26}$~s$^{-1}$) results in $Q({\rm H_2O})=(6.5\pm1.0)\times10^{26}$\,s$^{-1}$ at the epoch of our ALMA observations.  Modeling was carried out for a range of coma kinetic temperatures $T_{kin}=20$--70~K, based on the temperatures previously observed at radio and infrared wavelengths for comets in the distance range 1.7-2.4~au (Supplementary Table 1). 

Flux losses due to the filtering of large-scale coma structures by the interferometer were simulated by processing the (three-dimensional) synthetic coma images using the CASA {\tt simobserve} task, with the same hour angle, observation duration and array configuration parameters used for our science observations. The resulting synthetic visibilities were cleaned and imaged with the same parameters used for the observed cometary images.

The molecules of interest (X) were initially modeled assuming an isotropic, uniformly expanding coma, with outflow velocity ($v_{out}$) and production rate $Q({\rm X})$ allowed to vary as free parameters. The quality of the fits was monitored using the reduced chi-squared statistic ($\chi_r^2$), resulting in $\chi_r^2=1.03$ for our best fit to the CO interferometric data (Figure 2).  The best fit for HCN resulted in a surprisingly small value of $v_{out}=0.23\pm0.06$~km\,s$^{-1}$, so this fit was re-done, fixing the outflow velocity at the best-fitting value for CO ($v_{out}=0.47$~km\,s$^{-1}$; the CO autocorrelation data also gave $v_{out}=0.45\pm0.03$~km\,s$^{-1}$). This led to only a minor degradation of the HCN $\chi_r^2$ value, from 1.07 to 1.13, and is considered a more physically plausible scenario given typical coma outflow velocities $\sim0.5$-$0.7$~km\,s$^{-1}$ at $r_H=2$~au\cite{coc93,biv99,biv19}, with the assumption that HCN and CO are well mixed. The corresponding CO and HCN production rates are given in Table 1. The HCN emission is optically thin, so the derived production rate scales linearly with the assumed outflow velocity. An excellent fit to the HCN spectrum was also obtained using a conical jet model (see Figure 1), with $v_{out}=0.47$~km\,s$^{-1}$, and assuming all outgassing was confined to a conical region emanating from the nucleus, with its axis perpendicular to the line of sight, and a best-fitting cone opening angle of $96^{\circ}$. This resulted in a $13$\% increase in the derived HCN production rate, which is within the range of our observational uncertainties.

Our nominal model assumed a standard electron density scaling factor of $x_{n_e}=1.0$ and kinetic temperature $T_{kin}=$~50~K to obtain the results given in Table 1, but values as low as $x_{n_e}=0.2$ have been found to be more appropriate for some comets\cite{biv07,har10,biv19}. For $x_{n_e}=1.0$, the CO and HCN production rates are relatively insensitive to the assumed kinetic temperature, changing by less than 8\% across the range $T_{kin}=$~20--70~K. However, in the case of $x_{n_e}=0.2$, the HCN production rate is more sensitive to the value of $T_{kin}$ due to a reduction in electron collisions (and consequent drop in the $J=4$ level population), resulting in a value of $Q({\rm HCN})=1.2\times10^{24}$\,s$^{-1}$ at 20~K (compared with $7.0\times10^{23}$\,s$^{-1}$ for our standard model with $T_{kin}=50$~K; $x_{n_e}=1.0$). For comparison, we found $Q({\rm CO})=4.6\times10^{26}$\,s$^{-1}$ at $T_{kin}=20$~K. Consequently, in the regime of very low kinetic temperature and electron collision rates, 2I/Borisov's CO/HCN ratio could plausibly be as low as $370\pm80$, but this is still higher than the range of values (1.25--238) observed in previous Solar System comets (excluding the chemically peculiar outlier, C/2016 R2).

The slightly blue-shifted profile of the CO autocorrelation spectrum (Figure 3) provides evidence for asymmetric outgassing from 2I/Borisov's nucleus. Assuming hemispheric asymmetry along the Sun-comet vector, our best-fitting radiative transfer model for the autocorrelation (single dish) data indicates an expansion velocity of $v_{out}(1)=0.49$~km\,s$^{-1}$ in the sunward hemisphere, and $v_{out}(2)=0.26$~km\,s$^{-1}$ in the anti-sunward hemisphere, with an asymmetry factor (ratio of sunward to anti-sunward production rates) of 3.3. The isotropic model has a best-ftting $\chi^2_r=1.21$ (corresponding to $P = 0.02$, where $P$ is the probability that the difference between the model and observations is due to random noise), whereas the asymmetric model has $\chi^2_r=1.01$ and $P = 0.45$. Statistically, the 2-component, asymmetric model is therefore favoured. Our autocorrelation data are thus consistent with an enhanced CO sublimation rate on the illuminated side of the nucleus, which can be explained as a result of the increased Solar heating there. Since the CO $J=3-2$ emission is optically thin, the production rate for our asymmetric model is the same as that obtained using the standard, isotropic model ($Q({\rm CO})=(5.0\pm0.5)\times10^{26}$\,s$^{-1}$ at $T_{kin}=50$~K).

Three-sigma upper limits were obtained for the CS and CH$_3$OH production rates from the integrated spectral noise within a cometocentric velocity range $-0.8$ to $+0.6$~km\,s$^{-1}$ (based on the velocity range of detected flux in the CO autocorrelation spectrum). These correspond to mixing ratios of $<0.1$\% for CS and $<68$\% for CH$_3$OH, which are consistent with the values previously observed in other Solar System comets (with CS in the range 0.02--0.2\% and CH$_3$OH is in the range 0.6--6.2\%)\cite{boc18}.

\subsection*{Optical Observations of CN}

2I/Borisov was observed with the 60 cm TRAPPIST-South telescope located at La Silla Observatory, Chile\cite{jeh11} on December 15, 18 and 19 (UT). On each night a long (1500~s) exposure was obtained trough a CN (0-0) narrow-band filter\cite{far00} centered at 387~nm and various broad band filters (Johnson-Cousins B, V, Rc, Ic). TRAPPIST-South is equipped with a 2k$\times$2k CCD camera providing a field of view of $22'\times22'$ and the plate scale using a binning of $2\times2$ is $1.3''$/pixel. Data reduction followed standard procedures using frequently updated master bias, flat, dark frames and the flux calibration was performed with zero points obtained from regular standard star observations\cite{opi16,mou18}. To compute the CN production rate from the narrow-band images, we derived median radial brightness profiles for the CN and Rc images and used the continuum filter to remove the dust contamination. We then converted the flux to column density and fitted the profile with a Haser model at 10,000 km from the nucleus, using effective scale lengths from Cochran \& Schleicher\cite{coc93} and an outflow velocity of 0.5 km\,s$^{-1}$. The CN production rates obtained on December 15, 18 and 19, were respectively $(9.5\pm1.8)\times10^{23}$~s$^{-1}$, $(11.6\pm1.9)\times10^{23}$~s$^{-1}$, and $(9.2\pm1.7)\times10^{23}$~s$^{-1}$.

\section*{Acknowledgements}

We thank R. Simon for setting up the ALMA scheduling blocks, and D. Cruikshank for discussions on the composition of Kuiper belt objects. This work was supported by the National Science Foundation (under Grant No. AST-1614471), and by the Planetary Science Division Internal Scientist Funding Program through the Fundamental Laboratory Research (FLaRe) work package, as well as the NASA Astrobiology Institute through the Goddard Center for Astrobiology (proposal 13-13NAI7-0032). Part of this research was carried out at the Jet Propulsion Laboratory, California Institute of Technology, under a contract with the National Aeronautics and Space Administration. ALMA is a partnership of ESO, NSF (USA), NINS (Japan), NRC (Canada), NSC and ASIAA (Taiwan) and KASI (Republic of Korea), in cooperation with the Republic of Chile. The JAO is operated by ESO, AUI/NRAO and NAOJ.  The NRAO is a facility of the National Science Foundation operated under cooperative agreement by Associated Universities, Inc. TRAPPIST is a project funded by the Belgian Fonds (National) de la Recherche Scientifique (F.R.S.-FNRS) under grant FRFC 2.5.594.09.F. E.J. is a FNRS Senior Research Associate. N.X.R. was supported by the NASA Postdoctoral Program, administered by the Universities Space Research Association.

\includepdf[pages=-]{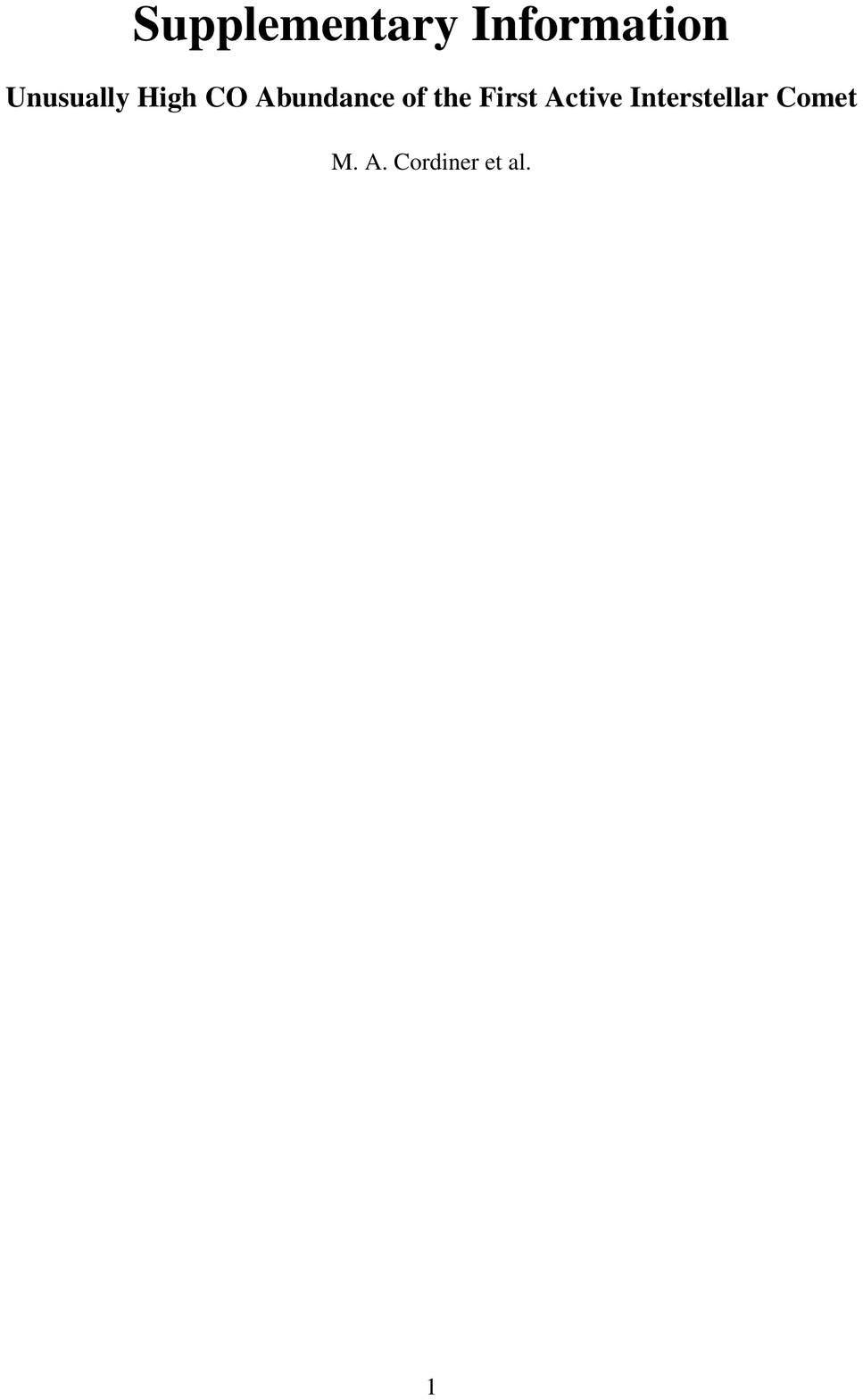}

\end{document}